\documentclass[conference]{IEEEtran}
\usepackage{adjustbox}
\usepackage{color}
\usepackage{booktabs}
\usepackage{multirow}
\usepackage{subfig}
\usepackage{comment}
\usepackage{csquotes}

\pdfoutput=1

\ifCLASSINFOpdf
\else
   \usepackage[dvips]{graphicx}
\fi
\usepackage{amsmath}
\usepackage{url}

\begin{document}
%
\title{Long-range Auto-correlations in Limit Order Book Markets: Inter- and Cross-event Analysis}



%
\author{\IEEEauthorblockN{Martin Magris\IEEEauthorrefmark{1}$^{,1}$,
Jiyeong Kim\IEEEauthorrefmark{2},
Esa R\"as\"anen\IEEEauthorrefmark{2}, 
Juho Kanniainen\IEEEauthorrefmark{1}
\IEEEauthorblockA{\IEEEauthorrefmark{1}Laboratory of Industrial and Information Management, Tampere University of Technology, Tampere, Finland}
\IEEEauthorblockA{\IEEEauthorrefmark{2}Laboratory of Physics, Tampere University of Technology, Tampere, Finland}
$^1$E-mail: martin.magris@tut.fi}}


\maketitle
\begin{abstract}
Long-range correlation in financial time series reflects the complex dynamics of the stock markets driven by algorithms and human decisions. Our analysis exploits ultra-high frequency order book data from NASDAQ Nordic over a period of three years to numerically estimate the power-law scaling exponents using detrended fluctuation analysis (DFA). We address inter-event durations (order to order, trade to trade, cancel to cancel) as well as cross-event durations (time from order submission to its trade or cancel). We find strong evidence of long-range correlation, which is consistent across different stocks and variables. However, given the crossovers in the DFA fluctuation functions, our results indicate that the long-range correlation in inter-event durations becomes stronger over a longer time scale, i.e., when moving from a range of hours to days and further to months. We also observe interesting associations between the scaling exponent and a number of economic variables, in particular, in the inter-trade time series.
\end{abstract}


%
\IEEEpeerreviewmaketitle

\section{Introduction}
Many natural and economic time series exhibit long-range power decaying correlations. 
The dynamics of the times series depicting a complex system is often characterized by \textit{scaling laws}, over a continuous range of \textit{time scales} and frequencies \cite{kantelhardt2012fractal}. A system characterized by self-similar structures over different time-scales is called a \textit{fractal}.
Financial and economic systems are highly complex and stochastic, characterized by numerous degrees of freedom and highly susceptible to exogenous factors.
This complexity emerges as time-dependent properties (e.g., \textit{trends}), or more generally, non-stationariety in the time-series. A reliable method for detecting long-range correlations, known as \textit{deterended fluctuation analysis} (DFA), which is robust for non-stationariety, has been developed in \cite{peng1994mosaic}. Because of its simplicity and wide applicability, DFA has been extensively used in the analyses of the long-range correlations in natural, social and economic data (see section \ref{sec:fin_references}).
However, in the existing literature DFA with intra-day high-frequency financial data has not been extensive.
The closest works related to our research are, among others, \cite{ivanov2004common, yuen2005impact, jiang2009detrended}, which studied the long memory and multifractal nature of inter-trade durations, and \cite{gu2014empirical}, which analyzed the long-range properties of inter-cancel durations and characterized their distribution.
Set apart from previous studies, we do not limit our analysis to one type of order book events. We compare the correlation properties in each type of order book event, i.e., order, trade or cancellation, and make cross-analyses between different events (e.g. order submission and its cancellation) as well. 

In this paper, we study the fractal properties of order flow data over different time scales, providing new insights into understanding the complex dynamics of the order book, also in relation to selected economic variables. In order to do so, we utilize the ultra-high frequency order book data, with five securities traded in NASDAQ Nordic over three years. We apply DFA on inter-event and cross-event time intervals of all message types (order submissions, transactions, and cancellations); this has not been done in the extant literature, even though message types are likely be interconnected. Our main finding is that fractal properties are ubiquitous in the time series of the order book, but of complex nature. The scaling properties show crossovers and peculiar relationships with a number of variables of economic interest (such as mean inter-event duration, daily return and volatility).

\section{Method and data}\label{sec:data_and_methods}
 
\subsection{Detrended fluctuation analysis} \label{subsec:DFA_algorithm}

Due to the non-stationary nature of financial time series, conventional methods, e.g., Hurst's rescaled range analysis \cite{hurst1951long}, can lead to a false detection of long-range correlations (see e.g., \cite{bryce2012revisiting}), as they assume stationary time series. Detrended fluctuation analysis (DFA), originally introduced in \cite{peng1994mosaic}, incorporates a detrending scheme in the fluctuation analysis, making the algorithm robust to non-stationarities like trends. Therefore, DFA has been established as a reliable method to analyze long-range correlation in financial time series.
We provide a brief description of DFA as it is our main analysis method. A detailed and thorough description of the algorithm can be found in \cite{kantelhardt2001detecting}.

For a time series of length $N$, whose observations are $\left\lbrace x_t \right\rbrace _{t=1,...,N}$, the DFA procedure can be summarized in four steps:
\begin{itemize}
\item[i.] We define the profile of the time series by taking an integrated sum of the series:
\begin{equation*}
y\left( k \right) = \sum_{t=1}^k \left(  x_t - \langle x\rangle \right)
\end{equation*}
The subtraction by the mean $\langle x\rangle$ sets the global mean to zero; however, it is not obligatory.

\item[ii.] The profile is divided into $N/s$ non-overlapping windows of equal length $s$. In each window, an $n$-degree polynomial approximation $y_{tr}$, representing local trend, is computed by a least-squares fit. In our analysis we use 1\textsuperscript{st} order DFA, in which a linear trend is eliminated from each window.

\item[iii.] We compute the variance of the residuals, or the detrended profile, $(y_m-y_{m,tr}), \, m=1,...,N/s$ for the $m$-th window and then average the variances. By taking the square root of the average variance, we obtain the DFA fluctuation $F$ as a function of window size $s$, as we repeat the procedure for all the window sizes:
\begin{equation}\label{eq:dfa}
F(s) =\sqrt{\frac{1}{N/s}\sum_{m=1}^{N/s} \left[ \frac{1}{s}\sum_{i=1}^{s}\left[y_m(i)-y_{tr, m}(i)\right]^2 \right]}
\end{equation}

\item[iv.] Since we have $F(s) \sim s^\alpha$ in presence of power-law scaling, we plot $F(s)$ against $s$ in log-log scale and calculate the slope of the linear fit in the log-log plot to obtain the \textit{scaling exponent} $\alpha$. Note that time series may require more than one scaling exponent to describe different correlation behaviors at different time scales. This \enquote{crossover} can be detected as change in slope in the log-log plot of $F(s)$ against $s$.

\end{itemize}
The scaling exponent $\alpha$ describes the nature of the correlation present in the data. White noise (uncorrelated signal) and Brownian noise are characterized by $\alpha=0.5$ and $\alpha=1.5$, respectively. Values $0.5 < \alpha < 1.5$ indicate long-term correlations, i.e., fractality, with $\alpha=1$ corresponding to perfect $1/f$ fractal behavior (pink noise). Values $\alpha<0.5$ correspond to anticorrelations \cite{bashan2008comparison, kantelhardt2012fractal}.

\subsection{DFA in financial literature}\label{sec:fin_references}
The nature of a power-law describing the long-range correlation implies self-similar patterns in the time series over a long period, which is of particular interest in economic and financial problems. 
The existence of long memory behavior in asset returns was first considered in \cite{mandelbrot1971can}, from which a rich literature on the fractal properties of financial time series followed. 
DFA analysis in finance is indeed ubiquitous and applied to various time series; we provide a number of examples.
In the earliest studies, \cite{mantegna1995scaling} showed that the scaling in the distribution of the S\&P 500 index can be described by a non-gaussian process and that the scaling exponent is constant over the period of six years analyzed. 
The first use of DFA on the evolution of currency rate exchange was proposed in \cite{vandewalle1997coherent}, where the authors found a close association between the scaling exponent and economic events, economic policies, and the information propagation among economic systems. These analyses have been expanded to a wider number of exchange rates in \cite{vandewalle1997detrended}. 
The use of the long-range correlation analysis to identify different phases in the evolution of financial markets and predict forthcoming changes (e.g. financial crashes) is disussed in, e.g., \cite{grech2004can,czarnecki2008comparison}.
Examples of long-range correlation analyses applied to financial returns can be found in, among the others, \cite{benbachir2011multifractal, carbone2004time, oswie2005multifractality, alvarez2008time}. 
Among the latest studies, \cite{thompson2016multifractal} compared the results implied by DFA with those from standard time series models, while \cite{tiwari2017multifractal} provides a very interesting application of DFA to investigate market efficiency of Dow Jones ETF.

Previous studies related to the present work include \cite{ivanov2004common}, which analyzed the distribution and fractal behavior of the inter-trade durations over a four-years period for thirty stocks listed at NYSE. Time intervals between consecutive trades were also considered in \cite{yuen2005impact} for stocks traded at NYSE and NASDAQ, finding that power-law correlations in inter-trade times are influenced by the market structure and coupled with the power-law correlations of absolute returns and volatility.
Inter-trade times for ultra-high frequency order book data were analyzed in \cite{jiang2008scaling} and \cite{jiang2009detrended}; in the latter work, strong evidence of crossovers between two different power-scaling regimes from 23 stocks was found. Fractal properties of inter-cancellations duration have been analyzed for 18 stocks in Shenzhen exchange in \cite{gu2014empirical}, using three variants of the standard DFA.

The above mentioned studies, though dealing with a large number of stocks, consider a limited number of variables, e.g., inter-trade or inter-cancellation times only. Joint analyses of the fractal structure of several variables (as proposed in \cite{gu2014empirical}) are definitely interesting and need to be expanded to a larger number of time series.
The full order book data contain orders, trades, cancellations and cross-events for each day down to ultra-short time scales, thus allowing a wide perspective to the dynamics, especially when combined with standard economic variables (e.g. daily return or volatility).

\subsection{Data}
We processed the raw ITCH order book flow and reconstructed the order book for a selected number of securities; for a detailed description of the procedure, we refer to \cite{ntakaris2017benchmark}. The ITCH flow contains the full information about all the events related to a particular security;  therefore, our order book data is complete and has accurate timestamps with millisecond precision. 
For our analysis, we have selected five stocks from NASDAQ Nordic (see Table \ref{tab:stock_description}), based on their liquidity, which directly determines the number of events in the order book. 
More information on NASDAQ Nordic order book market can be found, e.g., in \cite{siikanen2017limit,siikanen2017drives}.
We consider 752 trading days ranging from June 1, 2010 to May 31, 2013. In order to avoid bias in the data, non-regular trading hours are excluded as well as the events occurred in the first and last 30 minutes of the trading day, thus considering the time window spanning from 7:30 a.m. to 3:00 p.m. (2:30 p.m. for the stock traded in Copenhagen). The variables we use in our analysis are listed in Table \ref{tab:var_description}. We also differentiate the variables based on the side of the order book, while focusing on the events occurring at the best levels only.

\begin{table*}[h!]
\centering
\caption{List of five stocks of our selection at NASDAQ Nordic. The average number of records in the order book for each variable is computed over 752 trading days and over bid and ask sides.}
	\label{tab:stock_description}
    \begin{tabular}{llllccccc}
    \multicolumn{1}{c}{\multirow{2}[3]{*}{\textbf{ID}}} & \multicolumn{1}{c}{\multirow{2}[3]{*}{\textbf{ISIN code}}} & \multicolumn{1}{c}{\multirow{2}[3]{*}{\textbf{Company}}} & \multicolumn{1}{c}{\multirow{2}[3]{*}{\textbf{Exchange}}} & \multicolumn{5}{c}{\textbf{Avg. number of records}} \\
\cmidrule{5-9}          &       &       &       & or-or & tr-tr & ca-ca & or-tr & or-ca \\
    \midrule
    DK\textsubscript{1}   & DK0010268606 & Vestas Wind Systems & Copenhagen & 4295  & 1730  & 2943  & 1020  & 3276 \\
    FI\textsubscript{1}   & FI0009005318 & Nokian Renkaat Oyj & Helsinki & 6405  & 1462  & 4812  & 903   & 5503 \\
    FI\textsubscript{2}   & FI0009007835 & Metso Oyj & Helsinki & 7480  & 1698  & 5628  & 1029  & 6452 \\
    SE\textsubscript{1}   & SE0000101032 & Atlas Copco A & Stockholm & 13031 & 2499  & 12330 & 1530  & 11502 \\
    SE\textsubscript{2}   & SE0000115446 & Volvo B & Stockholm & 15851 & 3883  & 14327 & 2406  & 13446 \\
    \bottomrule
    \end{tabular}%
\end{table*}%

\begin{table}[h!] 
\centering
    \caption{Time-variables and their description. For the cancel-cancel durations, only consecutive cancellations \textit{occurring at the best 	level} are taken into account.}
    \label{tab:var_description}
    \begin{tabular}{cl}
    \textbf{Variable} & \multicolumn{1}{c}{\textbf{Description}} \\
    \midrule
    or-or & Time to the next order (inter-order duration) \\
    tr-tr & Time to the next trade (inter-trade duration) \\
    ca-ca & Time to the next cancellation (inter-cancel duration) \\
    \multirow{2}[0]{*}{or-tr} & Lifetime of orders that led to a trade \\
          & (time from order submission to its trade) \\
    \multirow{2}[1]{*}{or-ca} & Lifetime of orders that have been canceled \\
          & (time from order submission to its cancellation) \\
    \bottomrule
    \end{tabular}%
\end{table}%

\begin{figure}[!h]
\centering
\includegraphics[width=0.95\linewidth]{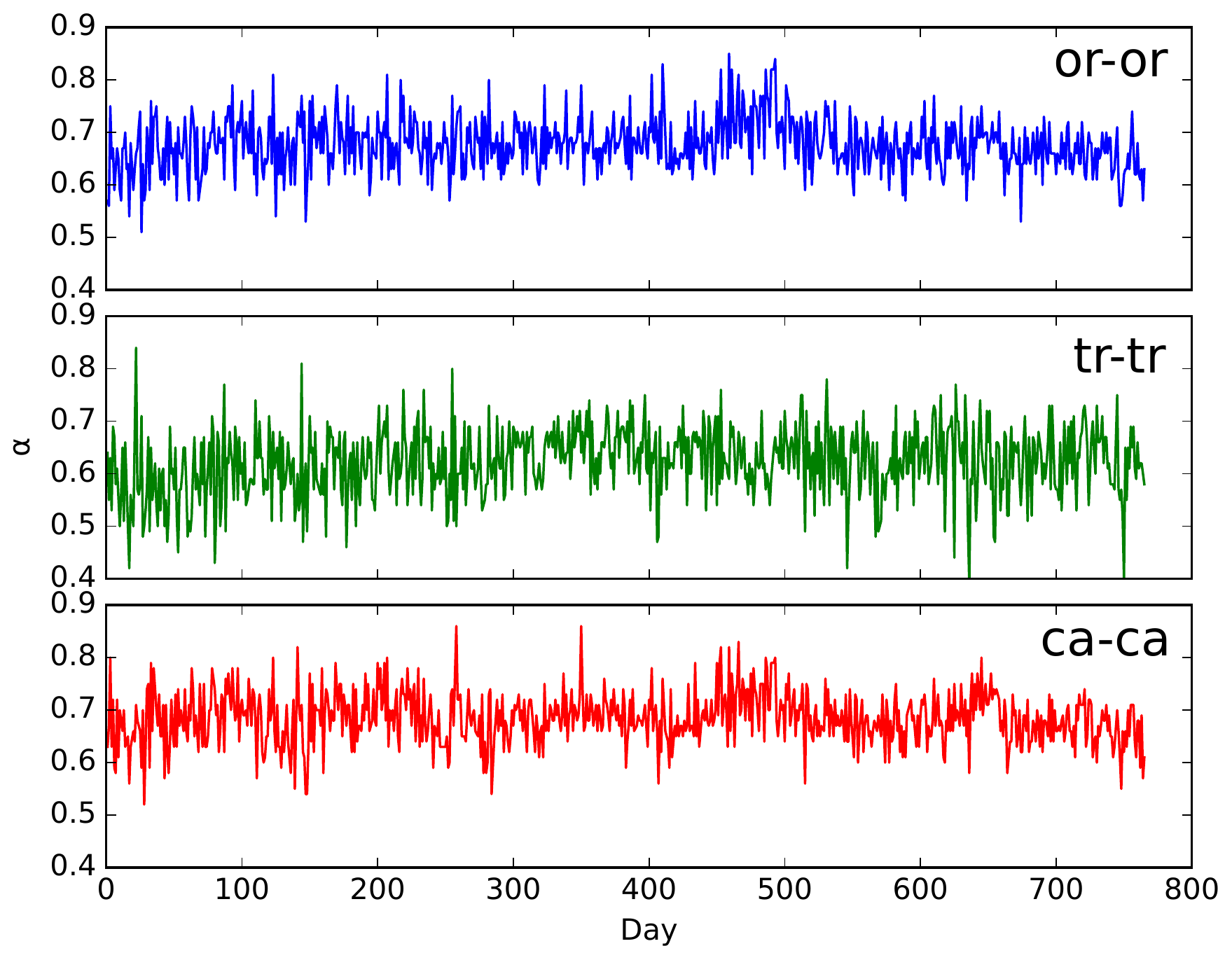}%
\label{fig_first_case}
\caption{Series of $\alpha$ values computed for inter-events durations for the bid side of stock FI\textsubscript{1}.}
\label{fig:aplhas_over_days}
\end{figure}

\section{Results and Discussion}
\subsection{Estimation of the power-law scaling exponent $\alpha$}

Using the DFA method described in section \ref{subsec:DFA_algorithm}, we compute the power-law scaling exponent $\alpha$ for each variable (Table \ref{tab:var_description}) that is recorded on a trading day. Fig. \ref{fig:aplhas_over_days} shows a sample series of $\alpha$ values over 752 trading days. The mean and standard deviation of the $\alpha$ values over all the available days are found in Table \ref{tab:mainResults}.

Across the multiple stocks we analyzed, the inter-event (or-or, tr-tr, and ca-ca) durations have the mean $\alpha$ values around 0.65, which are well above the threshold of random noise ($\alpha\approx0.5$), thus indicating the presence of (weak) long-range correlations in the waiting times within a day. The cross-events variables (or-tr and or-ca) also yield similar results, but with clearly larger day-to-day variance around the mean $\alpha$, compared to that of inter-event variables. 

The remarkable consistency of the results between the ask and bid sides suggests a symmetry of the correlation properties between the sides. 
This may imply that the way trading algorithms are implemented to look upon the past events on the bid and ask sides is very similar.

While mean $\alpha$ values are noticeably consistent between the stocks within a single exchange, there is slight heterogeneity in the mean alphas between the exchanges, especially for the cross-events variables. This may suggest that there are exchange-specific differences in the nature of the long-range correlation of stock market, since, e.g., not all the market participants (at NASDAQ Nordic) may trade at multiple exchanges.

\begin{table}[h!]
	\centering   
	\centering 
    \caption{Means and standard deviations of the daily scaling exponent $\alpha$, for different stocks and order book sides.}
    	
    \begin{tabular}{rccc}
    \multicolumn{1}{c}{\multirow{2}[3]{*}{\textbf{Stock}}} & \multirow{2}[3]{*}{\textbf{Variable}} & \textbf{Ask} & \textbf{Bid} \\
	\cmidrule{3-4}          &       & \textbf{Mean $\alpha$} & \textbf{Mean $\alpha$} \\
    \midrule
          & or-or & 0.68 $\pm$0.047 & 0.68 $\pm$0.050 \\
    \multicolumn{1}{c}{DK\textsubscript{1} } & tr-tr & 0.65 $\pm$0.066 & 0.65 $\pm$0.060 \\
    \multicolumn{1}{c}{Vestas Wind System} & ca-ca & 0.69 $\pm$0.057 & 0.68 $\pm$0.055 \\
          & or-tr & 0.57 $\pm$0.090 & 0.57 $\pm$0.098 \\
          & or-ca & 0.58 $\pm$0.104 & 0.59 $\pm$0.104 \\
    \midrule
          & or-or & 0.68 $\pm$0.050 & 0.68 $\pm$0.050 \\
    \multicolumn{1}{c}{FI\textsubscript{1}} & tr-tr & 0.63 $\pm$0.064 & 0.62 $\pm$0.065 \\
    \multicolumn{1}{c}{Nokian Renkaat Oyj} & ca-ca & 0.69 $\pm$0.051 & 0.68 $\pm$0.050 \\
          & or-tr & 0.58 $\pm$0.112 & 0.58 $\pm$0.114 \\
          & or-ca & 0.63 $\pm$0.077 & 0.63 $\pm$0.076 \\
    \midrule
          & or-or & 0.68 $\pm$0.046 & 0.68 $\pm$0.045 \\
    \multicolumn{1}{c}{FI\textsubscript{2}} & tr-tr & 0.63 $\pm$0.065 & 0.63 $\pm$0.064 \\
    \multicolumn{1}{c}{Metso Oyj} & ca-ca & 0.69 $\pm$0.047 & 0.69 $\pm$0.046 \\
          & or-tr & 0.58 $\pm$0.094 & 0.57 $\pm$0.091 \\
          & or-ca & 0.64 $\pm$0.082 & 0.63 $\pm$0.084 \\
    \midrule
          & or-or & 0.68 $\pm$0.039 & 0.68 $\pm$0.041 \\
    \multicolumn{1}{c}{SE\textsubscript{1}} & tr-tr & 0.65 $\pm$0.060 & 0.65 $\pm$0.060 \\
    \multicolumn{1}{c}{Atlas Copco A} & ca-ca & 0.68 $\pm$0.040 & 0.68 $\pm$0.041 \\
          & or-tr & 0.64 $\pm$0.098 & 0.64 $\pm$0.095 \\
          & or-ca & 0.71 $\pm$0.067 & 0.71 $\pm$0.069 \\
    \midrule
          & or-or & 0.69 $\pm$0.038 & 0.69 $\pm$0.038 \\
    \multicolumn{1}{c}{SE\textsubscript{2}} & tr-tr & 0.66 $\pm$0.050 & 0.66 $\pm$0.050 \\
    \multicolumn{1}{c}{Volvo B} & ca-ca & 0.68 $\pm$0.043 & 0.68 $\pm$0.045 \\
          & or-tr & 0.64 $\pm$0.088 & 0.64 $\pm$0.090 \\
          & or-ca & 0.71 $\pm$0.065 & 0.71 $\pm$0.064 \\
    \bottomrule
    \end{tabular}%
  \label{tab:mainResults}%
\end{table}%

\subsection{Crossovers in correlation behaviors}

\begin{figure}[!h]
\centering
\includegraphics[width=3in]{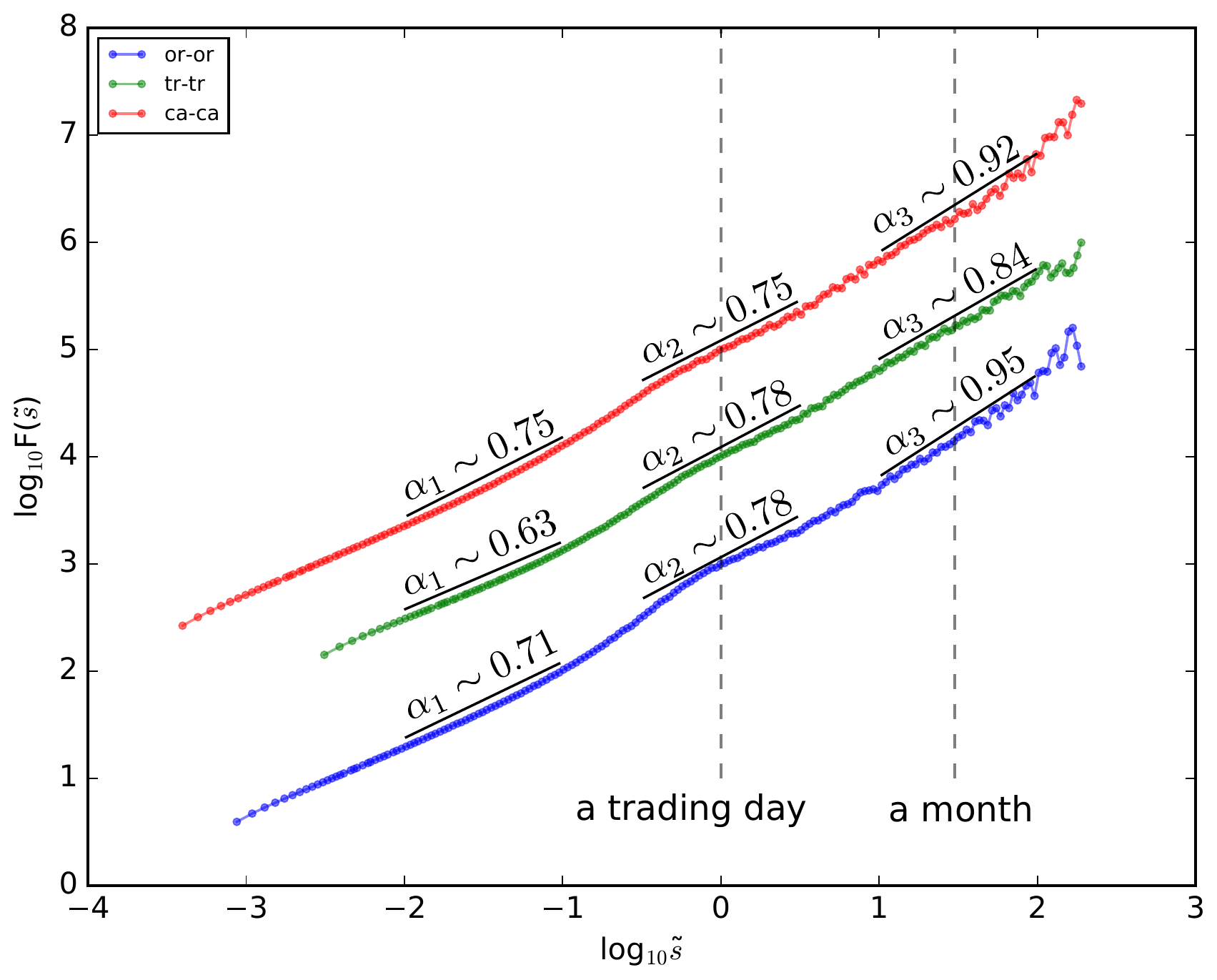}%
\label{fig_first_case}
\caption{Log-log plot of fluctuation \eqref{eq:dfa} as a function of normalized scale $\tilde{s}$ for the inter-event duration series on the bid side for the stock SE\textsubscript{1}. For each series we denote the local regressions and the corresponding slopes, i.e., $\alpha$.}
\label{fig:SE1_ask}
\end{figure}

By aggregating the daily recordings of a variable over all the available days, we observe crossover phenomena, i.e., changes in the scaling behavior at different time scales. Such crossovers in inter-event durations have been reported in earlier studies, e.g., \cite{ivanov2004common,tiwari2017multifractal}, in which two scaling exponents characterizing short- and long-range correlation are studied. In our analysis, however, motivated by the long time span of 3 years and the accuracy up to millisecond precision of the ultra-high frequency data, we compute three scaling exponents, $\alpha_1, \alpha_2$, and $\alpha_3$, characterizing the correlation properties at intra-day, day, and month time scales respectively (Fig. \ref{fig:SE1_ask}). 

In DFA, the estimates of $\alpha_1, \alpha_2$, and $\alpha_3$ are obtained by calculating the slopes locally within relevant time scale ranges in the log-log plot of the fluctuation $F(\tilde{s})$ against $\tilde{s}$, as shown in Fig. \ref{fig:SE1_ask}. Note that we use $s$ -- the number of events -- as our time variable. The time scale $\tilde{s}$ is normalized by the average number of the events (order, trade, or cancel) during a trading day. Intra-day $\alpha_1$ is calculated in the range 0.003-0.1 ($\times$ average daily activity), $\alpha_2$ in the range 0.3-3 and $\alpha_3$ in the range 10-100. Dotted lines in Fig. \ref{fig:SE1_ask}, represent the time scales of a trading day ($\log_{10}1=0$) and a month ($\log_{10}30=1.48$) respectively. Coherent with \cite{ivanov2004common}, we also observe a \enquote{bump} around the time scale of a trading day, indicating a clear crossover.

\begin{figure*}[!h]
\begin{center}
\includegraphics[width=0.9\linewidth]{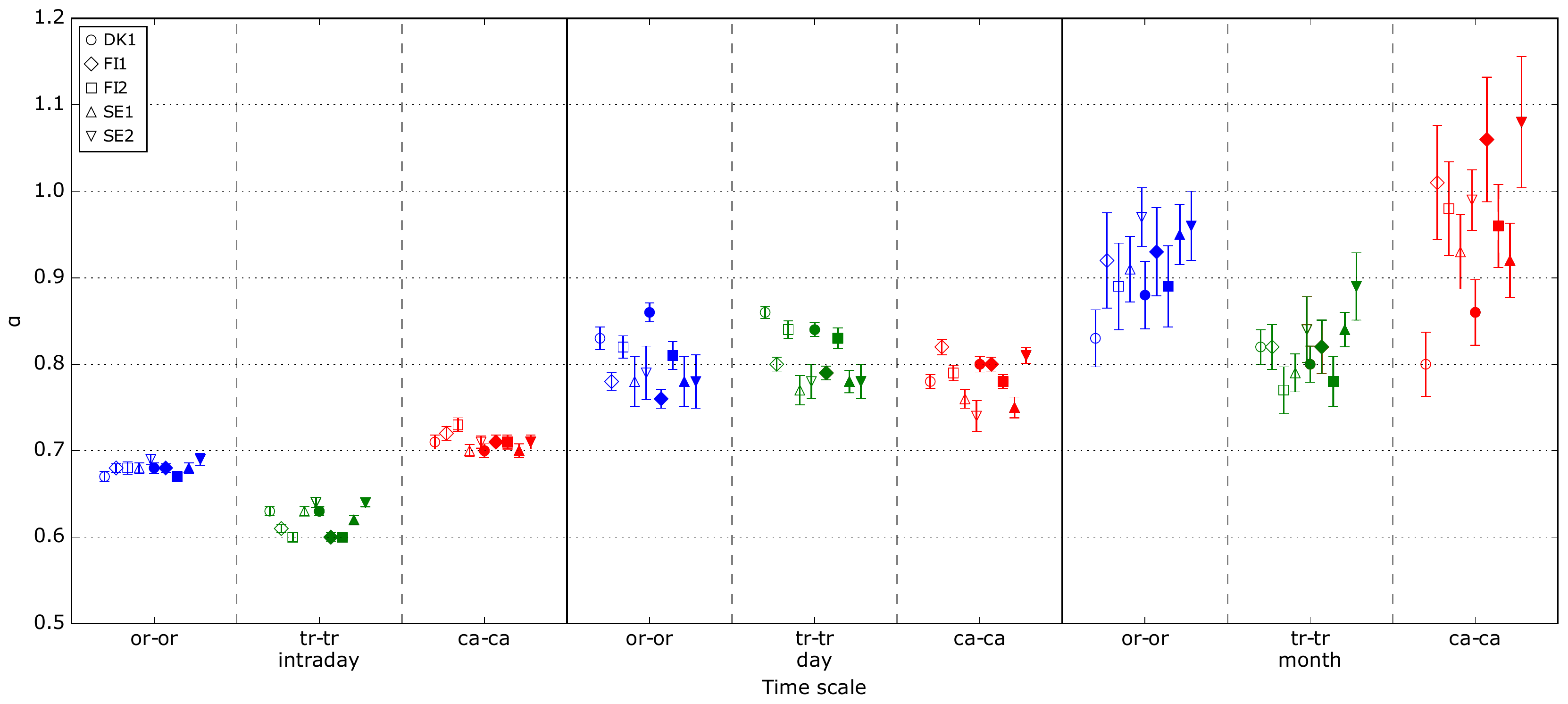}
\caption{Local $\alpha$ values computed at different time scales for each inter-event variable and stock. We report the 95\% confidence intervals. Filled (empty) markers represent bid (ask) side data.}
\label{fig:crossovers_vs_scale}
\end{center}
\end{figure*}

The final result is summarized in Fig. \ref{fig:crossovers_vs_scale}. For all inter-event variables, a significant increase in $\alpha$ is observed from intra-day to a day scale. Difference between $\alpha_2$ and $\alpha_3$ are not as prominent due to much bigger variance of $\alpha_3$; however, Fig. \ref{fig:crossovers_vs_scale} suggests that there is an increase in $\alpha$ from a day to a month scale for or-or and ca-ca durations. Increase in $\alpha$ at larger time scale signifies that the long-range correlation in inter-event durations become strong over longer time. This agrees with the economic theory (see e.g. \cite{fama1988dividend, campbell1997econometrics}) and supports the heterogeneous markets hypotheses of \cite{muller1993fractals}.

\subsection{Correlation between the scaling exponent and economic variables}
Utilizing the complete order book data available to us, we perform further analyses on the relationship between the power-law scaling exponent $\alpha$ and a selected number of market variables of interest. From financial point of view, it is interesting to understand whether there are some market variables that are closely correlated with the scaling behavior of the market flow.
We investigate the correlation between the mean intra-day $\alpha$ of inter-event durations and the following economic variables: i) avg. duration: mean inter-event duration of a trading day, ii) activity: number of orders, trades, or cancels during a trading day, i.e., length of the inter-event duration time series, iii) avg. quantity: mean quantity (volume) of orders, trades, or cancels in a trading day, iv) daily (log) return, v) volatility\footnote{As a proxy for the daily volatility, the realized variance (RV) is used. We implement the sub-sampling and averaging estimator of \cite{zhang2005tale}, estimating and averaging the RV over different sub-sampling grids of five minutes.}. The first four variables are directly related to the very same process, that generates the inter-event time variables, while daily return and volatility are determined by the (mid-) price process. 

\begin{figure}[!h]
\centering
\subfloat{\includegraphics[width=0.9\linewidth]{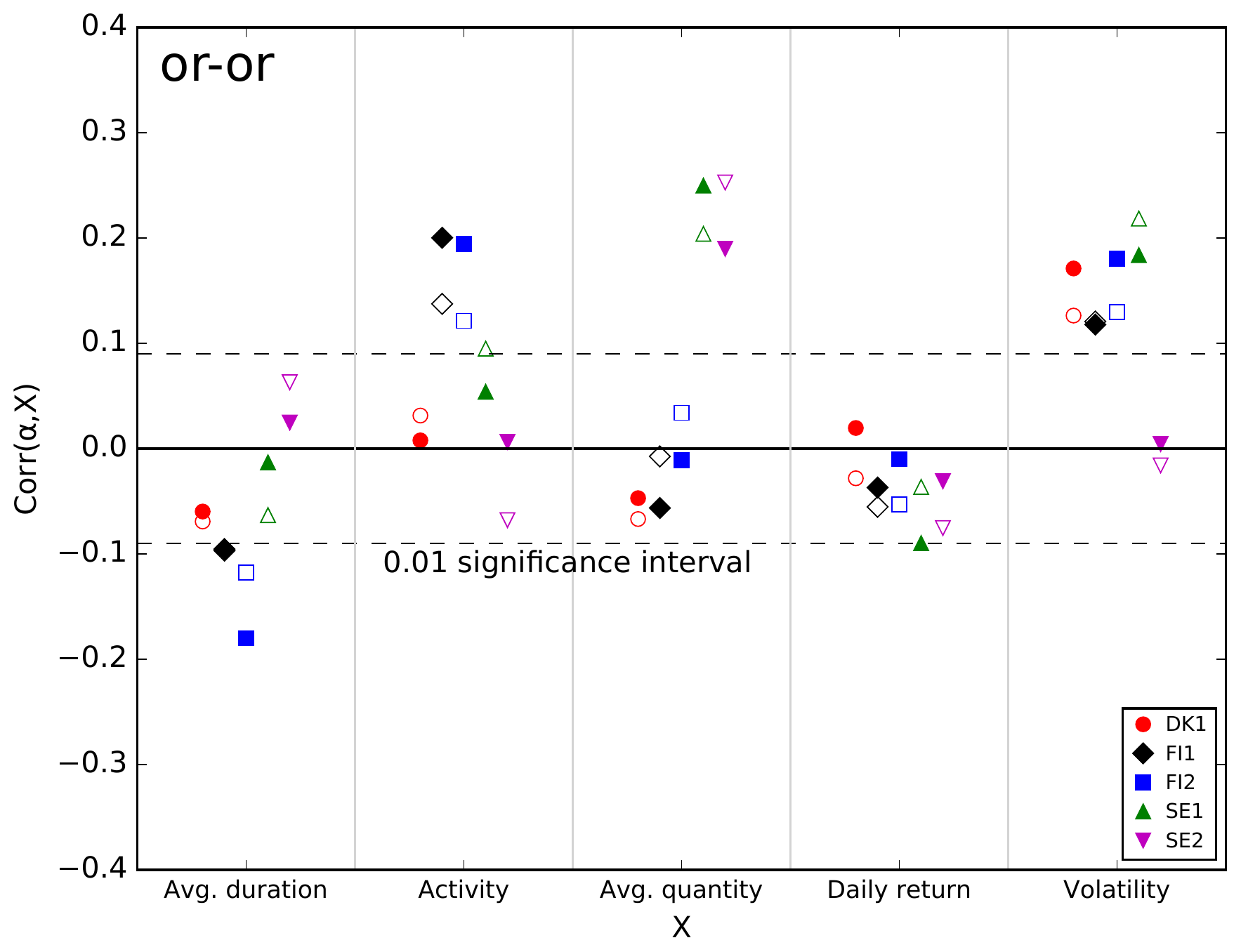}%
\label{fig:DFAalpha_correlation_or-or}}
\hfil
\subfloat{\includegraphics[width=0.9\linewidth]{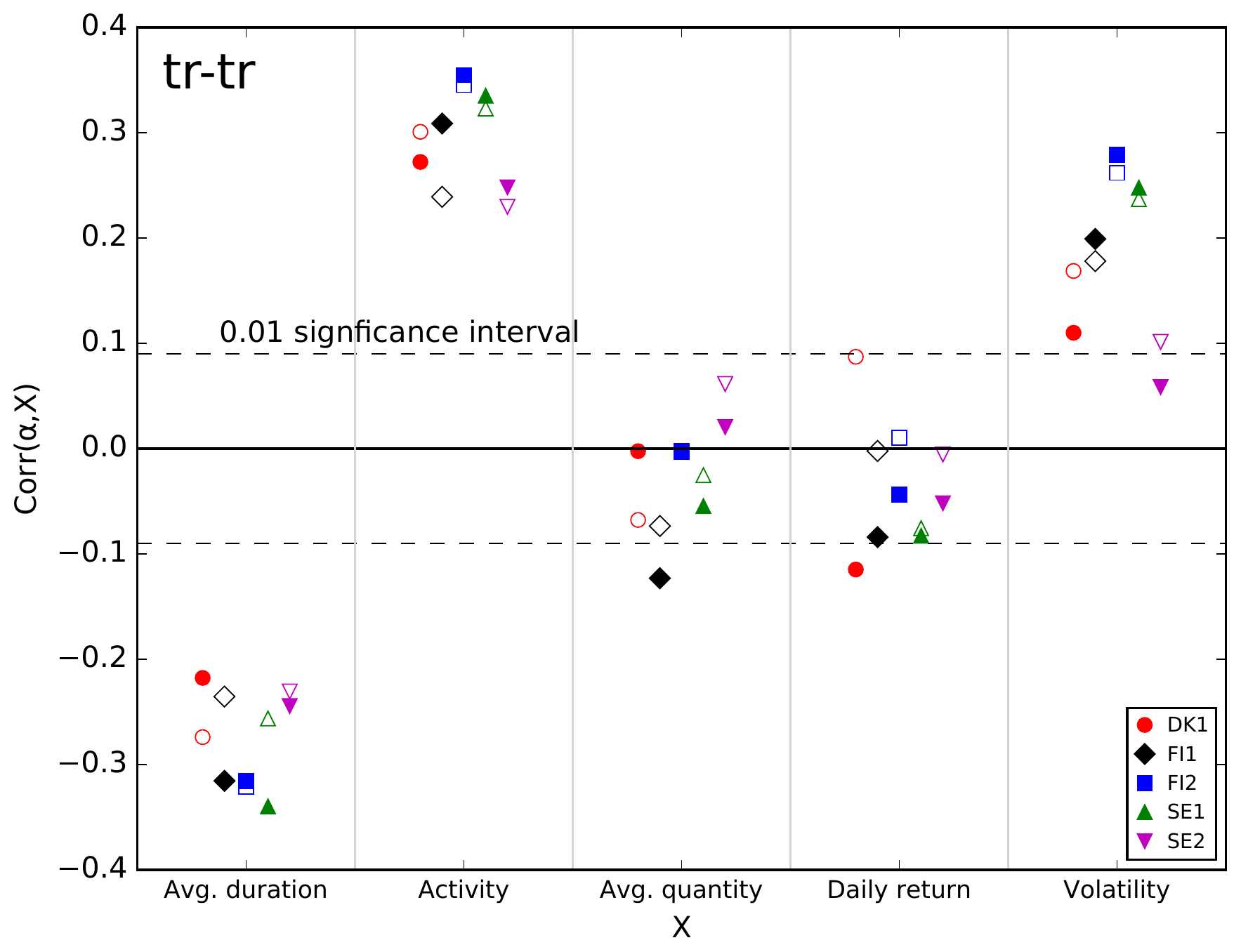}%
\label{fig:DFAalpha_correlation_tr-tr}}
\hfil
\subfloat{\includegraphics[width=0.9\linewidth]{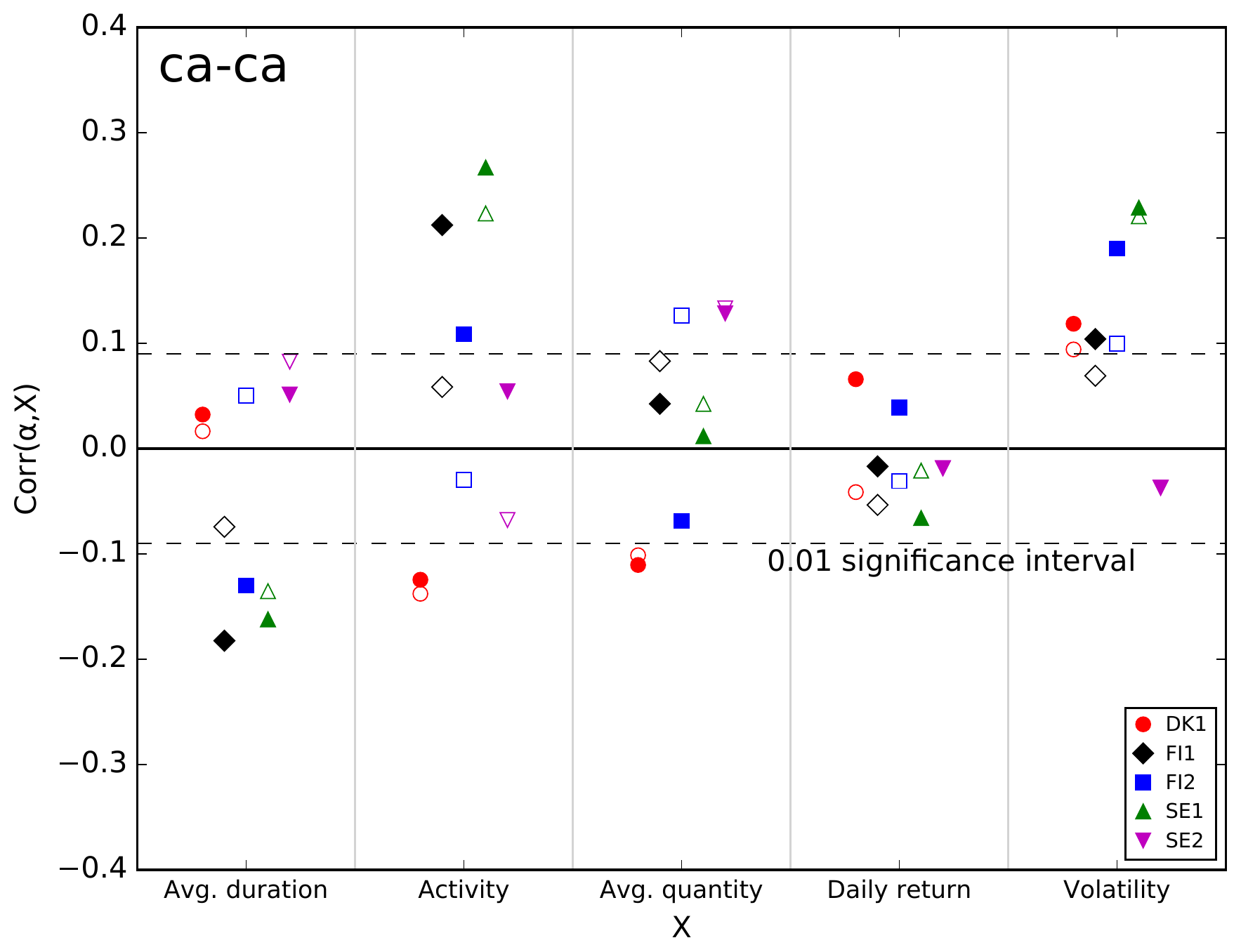}%
\label{fig:DFAalpha_correlation_ca-ca}}
\caption{Correlations coefficients between the daily $\alpha$ values and market variables, for the inter-event series. Filled and empty markers respectively denote bid and ask side data. Dashed lines mark the significance threshold, beyond which correlation is statistically significant at 99\%.}
\label{fig:crossovers_all}
\end{figure}

In Fig. \ref{fig:crossovers_all}, we present the results from the correlation analysis by plotting the (Pearson) correlation coefficient between $\alpha$ and $X$, the economic variables, as mentioned above. Stronger correlation is characterized by a larger magnitude of the correlation coefficient. Significance level of 0.01 is marked by dashed lines. 

For the time series of tr-tr durations, the correlation coefficients of all the stocks are clustered together for each variable. The clustering pattern among the stocks suggests that the correlation properties in tr-tr durations are not stock-specific, but may apply at a more general level. Consistency in the correlation measures between stocks within a same exchange is also not observed. On the other hand, the correlation coefficients of bid and ask sides are often found near to each other, confirming again the symmetry of the order book between the sides.

Average inter-trade duration and $\alpha$ are negatively correlated, while activity and volatility are positively correlated to $\alpha$.  Negative correlation between average inter-trade duration and $\alpha$ is in agreement with the previous study with NYSE and NASDAQ stocks \cite{yuen2005impact}. The significant correlation found between volatility and $\alpha$ is of financial interpretation. In periods of high volatility, trades tend to adjust their positions more often and accordingly become more reactive to other participants' submissions and cancellations. Furthermore, high volatility forces the participants to closely track the market evolution and make consequent decisions based on the market history from the past up to the most recent events, thus resulting in long-range dependence in tr-tr durations, i.e., larger $\alpha$. High volatility also infers high activity and shorter tr-tr duration; therefore, the positive and negative correlation respectively with $\alpha$ provides a cross-check in the reasoning. 

On the other hand, very weak (but not statistically significant) or no correlation is observed between $\alpha$ and average quantity or daily return. Indeed, market returns are known to be difficult to predict and either weakly dependent or not related to most of the economic variables \cite{cont2005long}. The lack of correlation indicates that the dynamics, which lead to long-range correlation in the inter-event durations, has little impact on the price evolution and its level at the end of the day. Similarly, our results also shows that the (daily mean) quantity of orders, trades, or cancels is mostly uncorrelated with time-related variables, as it has no significant impact on $\alpha$.

Interestingly, the clustering patterns in tr-tr duration is absent in or-or and ca-ca durations. The correlation coefficients of or-or and especially ca-ca are widely spread out, sometimes even across both positive and negative range. Therefore, it is more challenging to draw a strong conclusion from them.

\section{Conclusion}
In this paper we provide an extensive empirical analysis on the scaling behavior of the order book flow for selected stocks traded at NASDAQ Nordic. We perform DFA, a reliable analysis method for detecting long-range correlation in a time series, on the ultra-high frequency order book data with millisecond precision. We calculate the DFA scaling exponent $\alpha$, which characterizes the nature of the correlation present in the time series of inter-trade, -order, -cancel durations, yielding consistent results with previous works, e.g., \cite{yuen2005impact,jiang2009detrended,gu2014empirical}. We do not limit our analysis to inter-event durations, but extend the analysis to consider cross-events durations, such as lifetimes from order submission to its trade or cancellation. We find an outstanding consistency in the estimates of $\alpha$ values across different stocks and trading days, providing evidence that the time-related variables in the order book data are long-range correlated, i.e., fractal.

Taking the advantage of a complete and accurate order book data of 752 days, we investigate over different time scales, finding a significant change in correlation properties. This \enquote{crossover} phenomenon, also addressed in the past by, e.g.,  \cite{ivanov2004common,tiwari2017multifractal}, shows that the long-range correlation in intra-day inter-event durations becomes stronger at time scales beyond a day. The crossover analysis sheds light on the complexity of the market flow, as it takes more than one $\alpha$ to fully characterize the long-range correlation in the inter-event durations.

We also study the relationship between intra-day $\alpha$ and simple economic explanatory variables, such as mean inter-event duration, activity, mean quantity, daily return, and volatility. 
We find that, for tr-tr durations, the correlation coefficients of different stocks of both bid and ask sides form a cluster for each economic variable, while the clustering patterns are absent in or-or and ca-ca durations.
We explain from a financial point of view, the negative correlation between the average tr-tr duration over a day and $\alpha$, the positive correlation that activity and volatility have with $\alpha$, and the little to no correlation shown by average trade quantity and daily return with $\alpha$.

Our analysis reveals the complexity of the dynamics of stock markets, driven by automatic algorithms and human decision.
The work presented in this paper will be expanded to include a wider number of stocks to further investigate other variables that may be relevant to the long-range dependence in the order book flow, addressing the dependence of the scaling exponent on the time scale. We will consider not only time intervals but also include a fractal analysis for the volumes of orders, trades and cancellations. Moreover, in order to gain even better insights into the market dynamics, the present work will be enriched with more sophisticated analyses, i.e., implementing other statistical methods and variations of DFA (such as multifractal DFA and multiscale entropy analysis).

\section*{Acknowledgment}
The research leading to these results received funding from the Academy of Finland, the Finnish Academy of Science and Letters, and from the European Union's Horizon 2020 research and innovation program under Marie Sk{\l}odowska-Curie grant agreement No. 675044.



\bibliographystyle{IEEEtran}
\bibliography{IEEEabrv,biblio_file,biblio_file_redo}
%
%
%
\end{document}